\documentclass[10pt,twocolumn,letterpaper]{article}

\usepackage{./wss2026_template/cvpr}
\usepackage{times}
\usepackage{epsfig}
\usepackage{graphicx}
\usepackage{amsmath}
\usepackage{amssymb}
\usepackage{balance}
\usepackage{pgfplots}
\usepackage{booktabs}

\pgfplotsset{compat=1.18}
\usepgfplotslibrary{groupplots}

\usepackage[pagebackref=true,breaklinks=true,letterpaper=true,colorlinks,bookmarks=false]{hyperref}

\cvprfinalcopy 

\def\cvprPaperID{0009}

\newcommand{\ours}{\textbf{\textit{CounterFlow}}\xspace}

\ifcvprfinal\fi
\begin{document}

\title{CounterFlow: A Two-Phase Inference-Time Sampling for Counterfactual Video Foley Generation}

\author{
{\normalsize Gyubin Lee$^{1}$ \qquad
Junwon Lee$^{1}$ \qquad
Juhan Nam$^{1,2}$}\\[0.35em]
{\small
$^{1}$Kim Jaechul Graduate School of AI, KAIST, 
$^{2}$Graduate School of Cultural Technology, KAIST}\\[0.25em]
{\ttfamily\small \{gbstorm81,james39,juhan.nam\}@kaist.ac.kr}
}

\maketitle
\thispagestyle{empty}

\begin{abstract}
We investigate Counterfactual Video Foley Generation, which aims to adopt a sound-source identity that contradicts the visual evidence while remaining temporally synchronized to a silent video.
Existing Video\&Text-to-Audio (VT2A) models struggle with this, often remaining anchored to the visually implied sound source when video and text contents disagree. 
We present \ours, an inference-time dual-phase sampling scheme for pretrained flow-matching VT2A models. Phase~1 builds a video-derived temporal structure while suppressing the visually implied source; Phase~2 drops video conditioning to focus entirely on shaping audio timbre toward the target prompt. 
\ours substantially improves counterfactual Video Foley generation compared to naive negative prompting and state-of-the-art baselines. 
To evaluate replacement quality, we propose a metric leveraging a text-audio co-embedding space to measure both target-prompt evidence and residual visually implied source leakage. Video demonstrations and code are available at \url{https://gyubin-lee.github.io/counterflow-demo/}

\end{abstract}

\section{Introduction}
Foley sound production is fundamentally a controllable process: the timing of events may come from the video, but the sound that should be heard is often a designer's choice. A designer may therefore want to keep the motion of a visible event while changing its sound source, \eg, preserving a cat's motion while generating a lion roar, as in Fig.~\ref{fig:method_overview}. 

We define this task as \emph{Counterfactual Video Foley Generation}: given a silent video, a source text prompt describing the visible event, and a conflicting target text prompt, a generative model should output audio that preserves the video's temporal progression while reflecting the target sound source rather than the one implied by the video. This matters for creative sound design in film and game audio, where designers routinely replace a visible event without editing the video itself.

Existing methods struggle to address this specific challenge. While general VT2A models such as MMAudio~\cite{cheng2025mmaudio} and HunyuanVideo-Foley~\cite{shan2025hunyuanvideofoley} utilize both video and text conditioning, they remain optimized to generate the visually implied sound rather than replacing it with a conflicting target sound. Controllable Foley methods such as CAFA~\cite{benita2025cafa} and MultiFoley~\cite{chen2025multifoley} attempt to handle conflicting video and text, but they still exhibit a recurring limitation: the sampling trajectory often remains tied to the visually implied source, making reliable counterfactual video Foley generation difficult.
We therefore hypothesize that, in these pretrained flow-matching models, visual conditioning dominates the sampling trajectory, thereby weakening text control throughout the inference stages. 

We propose \ours, an inference-time technique that explicitly resolves the conflict between the input video and the target prompt. Driven by the intuition that early sampling steps primarily establish coarse event timing whereas later steps dictate counterfactual sound identity, we divide the sampling procedure into two distinct phases.

Our contributions are threefold: (1) \ours, an inference-time two-phase sampling method that separates video-guided temporal structure formation from subsequent target-sound injection; (2) a decomposed guidance design that suppresses the visually implied source during Phase~1 under conflicting video-text conditioning; and (3) a novel FLAM~\cite{wu2025flam}-based evaluation metric built on a text-audio co-embedding space to measure target sound fidelity and visually implied sound suppression simultaneously.

\section{Method}
We define the counterfactual video foley generation task as a Video\&Text-to-Audio (VT2A) problem where the generated audio follows the temporal dynamics of the video and matches the counterfactual sound identity described in the target text prompt, while suppressing the visually implied sound captured by the source prompt. The key technical challenge here is to \emph{suppress the visually implied sound identity within the video condition}, as video features often embed object-specific information.
Formally, given a silent video $V$, a target text $T_\mathrm{tar}$, and a source text $T_\mathrm{src}$ as input conditions, a flow-matching VT2A backbone predicts the velocity field $v_\theta(Z_t,c_{\mathrm{vid}},c_{\mathrm{txt}},t)$ in the audio latent space, where $c_\mathrm{vid}$ represents the video feature and  $c_{\mathrm{txt}}\!\in\!\{c_\mathrm{tar},\,c_\mathrm{src}\}$ denotes the text feature derived from either the target prompt, the source prompt produced by their respective pretrained encoders. The video and text conditions can each be independently disabled using their corresponding null embedding, $\emptyset_\mathrm{vid}$ and $\emptyset_\mathrm{txt}$.

\begin{figure*}[t]
\centering
\includegraphics[width=0.75\textwidth]{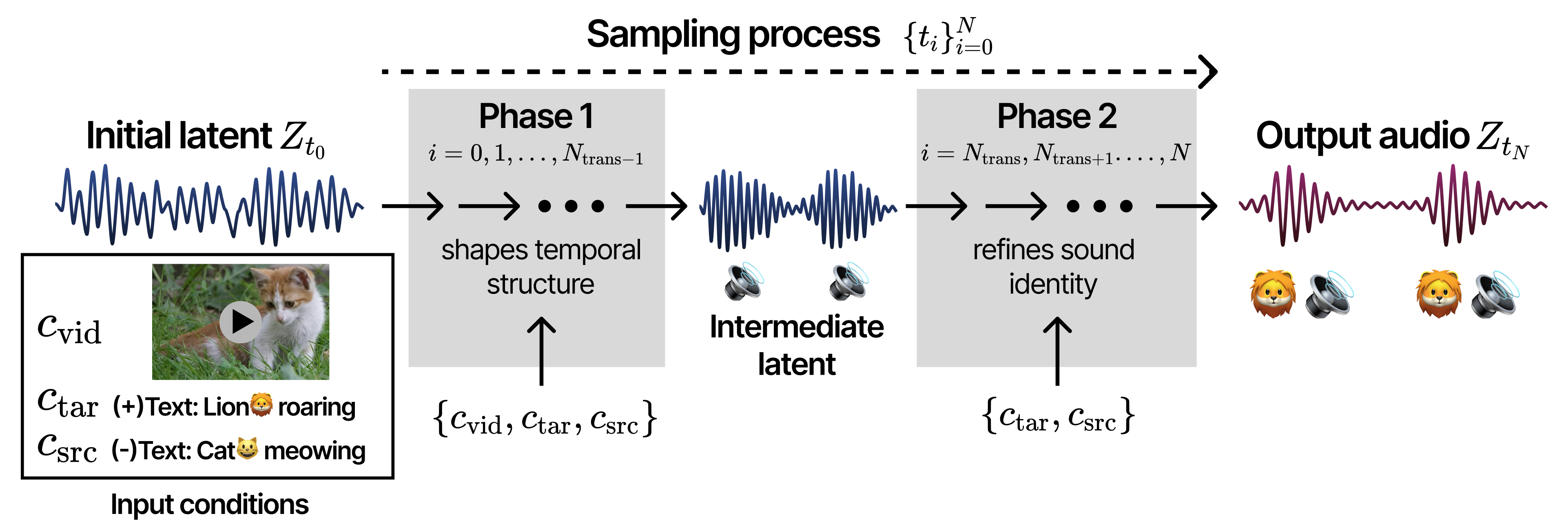}
\caption{\ours steers the sampling trajectory of a pretrained VT2A backbone at inference time without additional training. Phase~1 establishes a video-aligned temporal structure through decomposed guidance, while Phase~2 removes video conditioning and employs negative text prompting to refine the counterfactual sound identity within the established structure.}
\vspace{-1em}
\label{fig:method_overview}
\end{figure*}

We propose \ours, an efficient inference-time sampling method that operates without retraining VT2A backbones on new $(c_{\mathrm{vid}},c_{\mathrm{tar}},c_\mathrm{src})$ data.
The key idea is to split the flow-matching process into two separate phases that utilize distinct conditioning strategies, as shown in Fig.~\ref{fig:method_overview}.
This design is driven by two main intuitions. First, existing VT2A models often fail to follow the target prompt when a conceptually conflicting video condition dominates the sampling process. Second, it is well established that early sampling steps dictate the overall macro-structure (i.e., the temporal dynamics of the audio), while later steps refine the identity and details (i.e., the sound source and timbre) of the generated output ~\cite{manor2024zero}.
Therefore, we propose applying video conditioning only during the initial steps that contribute to the temporal structure, while leveraging negative prompting via the source text to maximize identity-level control over the target prompt.

During inference, the phase transition occurs at step $t_i=N_\mathrm{trans}$ within the ODE sampling timestep grid $\{t_i\}_{i=0}^{N}$. Let the initial audio latent be defined as $Z_{t_0}\sim\mathcal{N}(0,I)$ and $v_i(c_{\mathrm{vid}},c_{\mathrm{txt}})=v_\theta(Z_{t_i},c_{\mathrm{vid}},c_{\mathrm{txt}},t_i)$ represent the velocity field prediction at step $i$. 

In \textbf{Phase~1} ($i\in[0,N_\mathrm{trans})$), we maintain video conditioning active but replace the vanilla classifier-free guidance (CFG)~\cite{ho2022cfg}, formulated as $v_i(\emptyset_\mathrm{vid},\emptyset_\mathrm{txt})
+ w(v_i(c_\mathrm{vid},c_\mathrm{tar}) - v_i(\emptyset_\mathrm{vid},\emptyset_\mathrm{txt}))$, with our proposed decomposed guidance inspired by \cite{liu2022composable}:
\begin{equation}
\begin{aligned}
v_i^{(1)} =\; v_i(\emptyset_\mathrm{vid},\emptyset_\mathrm{txt})
&+ w_{\mathrm{vid}}\!\left(v_i(c_{\mathrm{vid}},\emptyset_\mathrm{txt}) - v_i(\emptyset_\mathrm{vid},\emptyset_\mathrm{txt})\right) \\
&+ w_{\mathrm{txt}}\!\left(v_i(\emptyset_\mathrm{vid},c_{\mathrm{tar}}) - v_i(\emptyset_\mathrm{vid},c_{\mathrm{src}})\right).
\end{aligned}
\label{eq:phase1}
\end{equation}
The second term builds a video-derived temporal structure, while the third explicitly promotes the target prompt and suppresses the source prompt. This decomposition prevents the model from predicting low-fidelity velocity field based on the conceptually conflicting conditions $v_i(c_\mathrm{vid},c_\mathrm{tar})$.
In \textbf{Phase~2} ($i\in[N_\mathrm{trans},N]$), the video conditioning is removed, and the intermediate state is refined using negative text prompting:
\begin{equation}
v_i^{(2)} = v_i(\emptyset_\mathrm{vid},\emptyset_\mathrm{txt})
 + w_{\mathrm{cfg}}\!\left(v_i(\emptyset_\mathrm{vid},c_{\mathrm{tar}})
 - v_i(\emptyset_\mathrm{vid},c_{\mathrm{src}})\right).
 \label{eq:phase2}
\end{equation}
The final latent $Z_{t_N}$ is decoded into a waveform via an audio VAE decoder.

\section{Experiments}
\noindent\textbf{Dataset.} We evaluate \ours on the two requirements of establishing a target identity while preserving video timing. To evaluate these capabilities, we use the VGGSound-Sparse Clean subset~\cite{iashin2024synchformer}, a clean single-source benchmark comprising 451 test videos across 12 unique sound source captions. For each video, we use the annotated caption as the source prompt and pair it with the other 11 captions as target prompts, resulting in 4,961 $(c_\mathrm{vid},c_\mathrm{tar},c_\mathrm{src})$ triplets. This setup enforces the conflict between the sounding object in the video and the sound identity specified by the target prompt.

\noindent\textbf{Implementation details.}
We use the pretrained MMAudio \texttt{large\_44k\_v2}~\cite{cheng2025mmaudio} as our backbone. We run \ours with deterministic Euler sampling for $N=25$ total sampling steps and generate 8-second outputs. We use $N_{\mathrm{trans}}=17$ for the phase transition, with guidance weights $w_{\mathrm{vid}}=3.0$, $w_{\mathrm{txt}}=5.0$, and $w_{\mathrm{cfg}}=4.5$.

\noindent\textbf{Baselines.}
We compare \ours against the state-of-the-art CAFA ~\cite{benita2025cafa}, our primary baseline for audio generation from conflicting video \& text, and the representative baseline ReWaS~\cite{jeong2025rewas}, which generates audio from a video-predicted sound energy curve and a text prompt describing the sound identity. We evaluate the first 8 seconds of CAFA's 10-second outputs to match our setting, while ReWaS metrics are computed on its native 5-second audios.

\noindent\textbf{Metrics.}
We report metrics across three criteria: FAD~\cite{kilgour2019fad} and Inception Score (IS)~\cite{salimans2016improvedgan} for overall audio quality and diversity, CLAP~\cite{wu2023large} for the identity relevance between the target prompt and the generated audio, and DeSync~\cite{iashin2024synchformer} for video-audio temporal alignment. 

To quantify the target-sound evidence against visually implied source leakage, we propose two novel metrics: $\Delta$FLAM and the positive-$\Delta$FLAM ratio.
Because FLAM~\cite{wu2025flam} provides frame-level sound event detection scores for individual sound events within simultaneous or sequential mixtures unlike clip-level scores, it allows us to compute a differential score comparing target and source evidence, an approach analogous to concurrent audio editing evaluations~\cite{chen2026audiochat}.
Let $A$ denote the generated audio and $P_{\mathrm{FLAM}}(c,A)=\max_l [p_{\mathrm{FLAM}}(c,A,l)]$ be the maximum frame-level probability across all frames $l$ for any text prompt $c$. 
We define
$\Delta\mathrm{FLAM}=P_{\mathrm{FLAM}}(c_{\mathrm{tar}},A)-P_{\mathrm{FLAM}}(c_{\mathrm{src}},A)$.
A higher $\Delta$FLAM indicates better counterfactual sound replacement. Crucially, this metric penalizes models that mistakenly generate sounds from both the target and source prompts, which standard CLAP scores often ignore. 
The positive-$\Delta$FLAM ratio measures the replacement success rate as 
$r_{>0}=\frac{1}{M}\sum_{m=1}^{M}\mathbf{1}\!\left[\Delta\mathrm{FLAM}^{(m)}>0\right]$,
representing the fraction of $M$ total clips where the target evidence exceeds the source evidence.


\section{Results}

\begin{table}[t]
\centering
\resizebox{\columnwidth}{!}{%
\begin{tabular}{lcccccc@{}}
\toprule
Method & FAD$\downarrow$ & IS$\uparrow$ & $\Delta$FLAM$\uparrow$ & (+)Ratio$\uparrow$ & CLAP$\uparrow$ & DeSync$\downarrow$ \\
\midrule
CAFA & 24.81 & 5.931 & 0.1289 & 0.8258 & 0.2371 & \textbf{0.5888} \\
CAFA + neg. & 31.46 & 7.606 & \underline{0.2573} & 0.8835 & 0.1801 & 0.6431 \\
ReWaS & 75.18 & 4.223 & 0.0560 & 0.6184 & 0.1084 & 1.078 \\
ReWaS + neg. & 79.52 & 4.703 & 0.1905 & 0.7130 & 0.0947 & 1.103 \\
\midrule
\ours & \underline{23.55} & \textbf{7.915} & \textbf{0.2641} & \textbf{0.9200} & \underline{0.2840} & 0.6695 \\
\hspace{0.5em} w/o P2 neg. & \textbf{23.29} & \underline{7.790} & 0.2373 & \underline{0.9170} & \textbf{0.2849} & \underline{0.6261} \\
\bottomrule
\end{tabular}%
}
\vspace{0.3em}
\caption{Main comparison under conflicting video-text control. 
`neg.' and P2 stand for negative prompting with $c_\mathrm{src}$ and Phase 2, respectively.
}
\vspace{-1em}
\label{tab:main_results}
\end{table}

\noindent\textbf{Quantitative analysis.} Table~\ref{tab:main_results} compares \ours against the state-of-the-art baselines.
Our method achieves the strongest overall counterfactual sound replacement performance and overall quality while remaining competitive in temporal alignment. 
External baselines struggle to replace the visually implied source sound with the counterfactual target sound. In contrast to \ours, they often generate sounds corresponding to both the target text and the video, which results in high CLAP scores and poor $\Delta$FLAM scores. This tendency is also evident in the qualitative analysis.
Adding negative prompting improves $\Delta$FLAM and its positive ratio, indicating that it helps suppress visually implied source identity from the video input. However, these gains are accompanied by degraded CLAP and DeSync scores, suggesting that applying negative prompting alone weakens the conditioning from both the target text and video, thereby deteriorating overall controllability.
The \ours \emph{w/o P2 neg.} configuration clarifies why the main method keeps negative source prompting in Phase~2. While removing this term slightly improves audio quality and temporal alignment, it compromises the FLAM-based replacement scores. This aligns with our methodological intuition: once Phase~1 has formed the temporal structure, maintaining negative source prompting active in Phase~2 helps facilitate target-sound injection without drifting back toward the visually implied source. Ultimately, our method avoids the severe performance degradation in FAD and CLAP scores observed in the other baselines, thanks to the decomposed guidance in Phase~1 (Eq.~\ref{eq:phase1}).

\begin{figure}[t]
\centering
\includegraphics[width=0.75\columnwidth]{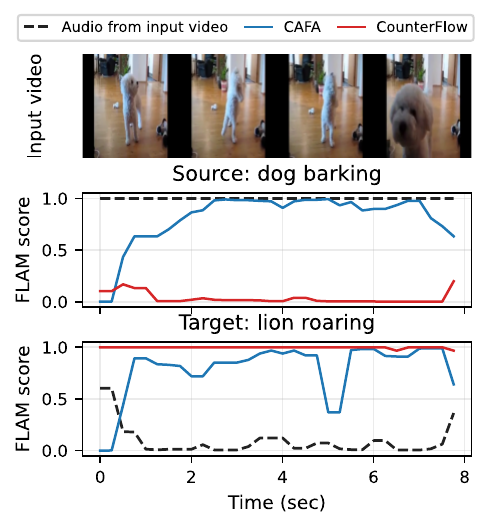}
\vspace{-0.5em}
\caption{FLAM visualization for a counterfactual video foley generation from \emph{dog barking} to \emph{lion roaring}. 
}
\vspace{-1em}
\label{fig:qualitative}
\end{figure}

\noindent\textbf{Qualitative analysis.}
Figure~\ref{fig:qualitative} illustrates how \ours achieves counterfactual video foley generation by suppressing visually implied source identity while injecting target-sound identity. Unlike CAFA, which remains tied to the original source evidence and consequently generates undesired visually implied sound events alongside the target sounds, \ours exhibits a clear contrast: the source-prompt FLAM score remains low, while the target-prompt FLAM score remains consistently high over the event duration. This confirms \ours's ability to replace the visually implied event with the target sound source while preserving the video's underlying temporal structure.

\noindent\textbf{Ablations.}
The ablations support three core design claims: decomposed Phase~1 guidance is necessary for counterfactual conditions, the temporal structure construction primarily occurs in Phase~1, and transition timing controls the replacement-temporal alignment trade-off. 
Table~\ref{tab:decomp_ablation} summarizes the quantitative comparisons. 
First, joint conditioning of the video and target prompts via vanilla CFG without decomposition in Phase~1 (Eq.~\ref{eq:phase1}) results in near-zero $\Delta$FLAM and CLAP scores. This indicates that the pretrained VT2A backbone prioritizes the video over the text condition when both are provided simultaneously, consequently preventing target-sound identity injection. Second, decomposition alone is insufficient; without explicit negative prompting using the source prompt in Phase~1 (i.e., $c_\mathrm{src}\rightarrow \emptyset_\mathrm{txt}$ in Eq.~\ref{eq:phase1}), the $\Delta$FLAM and CLAP scores deteriorate even with negative prompting active in Phase 2. This reveals that video features convey identity information on visual events, necessitating identity suppression via negative source prompting. Note that the lower DeSync scores observed in the simpler variants do not imply superior alignment, but rather reflect a failure to deviate from the original visually implied source.
Reversing the two phases (swapping Eq.~\ref{eq:phase1} and Eq.~\ref{eq:phase2}) leaves the $\Delta$FLAM and CLAP scores nearly unchanged but substantially worsens both FAD and DeSync. 
This provides the clearest support for our methodological intuition: early video-conditioned updates are crucial for deciding \emph{when} the sound should occur by developing a high-fidelity temporal structure, whereas later text-contrast updates are essential for deciding \emph{which} sound source should dominate.

\begin{table}[t]
\centering
\resizebox{0.9\columnwidth}{!}{%
\begin{tabular}{lcccc@{}}
\toprule
Method & FAD$\downarrow$ & $\Delta$FLAM$\uparrow$ & DeSync$\downarrow$ & CLAP$\uparrow$ \\
\midrule
\ours & 23.55 & 0.2641 & 0.6695 & 0.2840 \\
\hspace{0.5em} w/o P1 decomp. CFG & 24.36 & 0.0278 & 0.2390 & 0.0894 \\
\hspace{0.5em} w/o P1 neg. & 21.00 & 0.0534 & 0.4362 & 0.2608 \\
\hspace{0.5em} Phase swap (P1 $\leftrightarrow$ P2) & 52.33 & 0.2367 & 0.9989 & 0.2817 \\
\bottomrule
\end{tabular}%
}
\vspace{0.3em}
\caption{Ablation on decomposed text-video CFG and negative prompting, and swapping Phase 1 and 2. 
}
\label{tab:decomp_ablation}
\end{table}

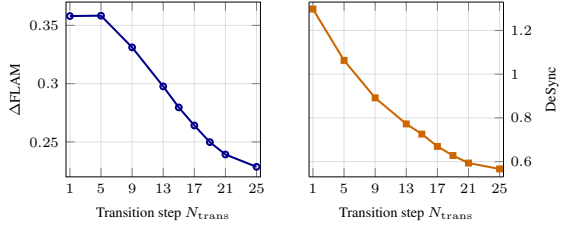
\begin{figure}[t]
\centering
\resizebox{0.9\columnwidth}{!}{\begin{tikzpicture}
\begin{groupplot}[
    group style={group size=2 by 1, horizontal sep=0.10\columnwidth},
    width=0.40\columnwidth,
    height=0.36\columnwidth,
    scale only axis,
    xmin=0.5, xmax=25.5,
    xtick={1,5,9,13,17,21,25},
    xlabel={Transition step $N_{\mathrm{trans}}$},
    xlabel style={font=\scriptsize},
    tick label style={font=\scriptsize},
    label style={font=\scriptsize},
    grid=major,
    grid style={draw=gray!25}
]
\nextgroupplot[
    ylabel={$\Delta$FLAM},
    ylabel style={font=\scriptsize, at={(axis description cs:-0.20,0.5)}},
    ymin=0.22, ymax=0.37,
    ytick={0.25,0.30,0.35}
]
\addplot[
    color=blue!55!black,
    line width=1.0pt,
    mark=o,
    mark options={scale=0.7, fill=white}
] coordinates {
    (1,0.3580)
    (5,0.3583)
    (9,0.3311)
    (13,0.2976)
    (15,0.2796)
    (17,0.2641)
    (19,0.2498)
    (21,0.2392)
    (25,0.2286)
};

\nextgroupplot[
    axis y line*=right,
    yticklabel pos=right,
    ylabel={DeSync},
    ylabel style={font=\scriptsize, at={(axis description cs:1.18,0.5)}},
    ymin=0.53, ymax=1.33,
    ytick={0.6,0.8,1.0,1.2},
    after end axis/.code={
        \draw[black] (rel axis cs:0,0) -- (rel axis cs:0,1);
    }
]
\addplot[
    color=orange!80!black,
    line width=1.0pt,
    mark=square*,
    mark options={scale=0.6}
] coordinates {
    (1,1.298)
    (5,1.063)
    (9,0.8917)
    (13,0.7723)
    (15,0.7259)
    (17,0.6695)
    (19,0.6276)
    (21,0.5932)
    (25,0.5663)
};
\end{groupplot}
\end{tikzpicture}}
\caption{Transition-step sweep on \ours. 
}
\vspace{-1em}
\label{fig:transition_sweep}
\end{figure}

Figure~\ref{fig:transition_sweep} demonstrates that transition timing dictates a trade-off between control over sound identity and temporal alignment, rather than yielding a monotonic improvement. As $N_{\mathrm{trans}}$ increases, the DeSync score improves, while $\Delta$FLAM steadily declines. An earlier switching, therefore, favors replacement, whereas a later transition favors alignment. We choose $N_{\mathrm{trans}}=17$ because it lies at the knee of the trade-off curve, preserving robust replacement performance while recovering most of the gains in temporal alignment.

\section{Conclusion}

We presented \ours, a two-phase inference strategy for counterfactual video Foley generation in pretrained VT2A models. By decoupling the sampling trajectory into temporal structure formation (Phase~1) and sound identity injection (Phase~2), our approach significantly improves replacement performance on the VGGSound-Sparse Clean subset without compromising audio quality or temporal alignment.
Despite these gains, \ours occasionally generates sound during silent intervals, indicating a limitation in strict temporal gating. Future work may address this by ensuring that generation is exclusively anchored to active visual cues via explicit training.
Furthermore, since our method is inherently model-agnostic, we plan to apply this framework to various VT2A backbones beyond MMAudio to further validate its generalizability.

\balance

\section*{Acknowledgments}
This work was partly supported by Institute for Information \& communications Technology Technology Planning \& Evaluation(IITP) grant funded by the Korea government(MSIT)(RS-2019-II190075, Artificial Intelligence Graduate School Support Program(KAIST)) and the National Research Foundation of Korea (NRF) grant funded by the Korea government (MSIT) (RS-2023-00222383).

{\small
\bibliographystyle{wss2026_template/ieee}
\bibliography{paper}
}

\end{document}